\documentclass[epj]{svjour}
\usepackage{amsmath,amssymb}
\usepackage{tensor}
\usepackage{etoolbox}
\usepackage{graphicx}
\usepackage{textgreek}
\usepackage{siunitx}
\DeclareSIUnit{\fm}{\femto\metre}
\let\Re\relax\DeclareMathOperator{\Re}{Re}

\newcommand*{\EquationComma}{\mathinner{,}}
\newcommand*{\EquationPeriod}{\mathinner{.}}
\DeclareMathOperator{\artanh}{artanh}
\newcommand*{\CMEPerNucleonPair}{\sqrt{s}_\text{NN}}
\newcommand*{\BottomQuark}{\text{b}}
\newcommand*{\Bottomium}{\text{\BottomQuark$\bar{\BottomQuark}$}}
\newcommand*{\UpsilonMeson}[1][]{\text{\textUpsilon\ifblank{#1}{}{(#1)}}}
\newcommand*{\chibMeson}[1][]{\text{\textchi\textsubscript{b}\ifblank{#1}{}{(#1)}}}
\newcommand*{\Proton}{\text{p}}
\newcommand*{\Lead}{\text{Pb}}
\newcommand*{\Pion}{\text{\textpi}}
\newcommand*{\CircleConstant}{\text{\textpi}}
\newcommand*{\EulersNumber}{\mathrm{e}}
\newcommand*{\Differential}[1]{\mathinner{\mathrm{d}#1}}
\newcommand*{\ImaginaryUnit}{\mathrm{i}}
\newcommand*{\Increment}{\text{\textDelta}}
\newcommand*{\Abs}[1]{\lvert{#1}\rvert}
\newcommand*{\Norm}[1]{\lVert{#1}\rVert}
\newcommand*{\PointChargeElectricField}{\vec{E}_q}
\newcommand*{\PointChargeMagneticField}{\vec{B}_q}
\newcommand*{\NucleusElectricField}{\vec{E}_\text{nuc}}
\newcommand*{\NucleusMagneticField}{\vec{B}_\text{nuc}}
\newcommand*{\TotalElectricField}{\vec{E}}
\newcommand*{\TotalMagneticField}{\vec{B}}
\newcommand*{\Conductivity}{\sigma}
\newcommand*{\Permittivity}{\varepsilon}
\newcommand*{\Permeability}{\mu}
\newcommand*{\EffectiveTime}{t_\text{eff}}
\newcommand*{\RadialDistance}{\varrho}
\newcommand*{\AzimuthalAngle}{\varphi}
\newcommand*{\ProperTime}{\tau}
\newcommand*{\LongitudinalRapidity}{y}
\newcommand*{\QGPFormationTemperature}{T_\text{crit}}
\newcommand*{\ThermalizationTime}{\ProperTime_\text{init}}
\newcommand*{\NumberOfCollisions}{N_\text{coll}}
\newcommand*{\NumberOfParticipants}{N_\text{part}}
\newcommand*{\Displacement}{r}
\newcommand*{\DisplacementRMS}{\Displacement_{\text{RMS},nl}}
\newcommand*{\NuclearModificationFactor}{R_{AA,nl}}
\newcommand*{\PreCascadeNuclearModificationFactor}{R^\text{QGP}_{AA,nl}}
\newcommand*{\TransverseMomentum}{p_\text{T}}
\newcommand*{\FormationTime}{\ProperTime_\text{F}}
\newcommand*{\ElectricDipoleAlignment}{\vartheta}
\begin{document}
\title{Electromagnetic field effects on \UpsilonMeson-meson dissociation in \Lead\Lead~collisions at LHC~energies}
\author{J. Hoelck \and G.Wolschin}
\institute{Institut f\"ur Theoretische Physik der Universit\"at Heidelberg, Philosophenweg 16, D-69120 Heidelberg, Germany}
\date{Received: date / Revised version: date} 
\abstract{
We investigate the effect of the electromagnetic field generated in relativistic heavy-ion collisions on the dissociation of \UpsilonMeson~mesons.
The electromagnetic field is calculated using a simple model which characterizes the emerging quark--gluon plasma~(QGP) by its conductivity only.
A numerical estimate of the field strength experienced by \UpsilonMeson~mesons embedded in the expanding QGP and its consequences on the \UpsilonMeson~dissociation is made.
The electromagnetic field effects prove to be negligible compared to the established strong-interaction suppression mechanisms.
%
\PACS{
	{25.75.-q}{Relativistic heavy-ion collisions}
	\and
	{24.10.Jv}{Relativistic models (nuclear reactions)}
	\and
	{25.75.Cj}{Heavy quark production in relativistic heavy-ion collisions}
	}
}
\maketitle
\section{Introduction}\enlargethispage{\baselineskip}
The suppression of quarkonia states in relativistic heavy-ion collisions as compared to \Proton\Proton~collisions at the same energy has been used as an indicator to assess the properties of the transient quark--gluon plasma.
Particularly clear evidence for the sequential suppression of individual quarkonia states has been obtained from \UpsilonMeson~data in \Lead\Lead~collisions at the CERN Large Hadron Collider (LHC)~\cite{CMS-2017-PLB}.
The suppression of the \UpsilonMeson[1S] and \UpsilonMeson[2S] states is found to rise strongly with increasing centrality, but it is essentially flat as a function of transverse momentum.

The large \BottomQuark-quark mass and tight \Bottomium~binding allow for a clean theoretical treatment of the \UpsilonMeson~meson in an expanding plasma of thermal gluons and light quarks.
Theoretical approaches such as \cite{SongHanKo-2012-PRC,ZhaoEmerickRapp-2013-NPA,ZhouXuZhuang-2014-NPA,KumarShuklaVogt-2015-PRC,KrouppaRyblewskiStrickland-2015-PRC,HoelckNendzigWolschin-2017-PRC} have been developed to relate the experimental findings to the quark--gluon plasma properties.
Our model encompasses screening of the real part of the quark--antiquark potential, collisional damping due to the imaginary part, direct dissociation by thermal gluons, and reduced feed-down in the final decay cascade~\cite{NendzigWolschin-2014-JPGNPP}.

A reasonable understanding of the centrality dependence of the spin-triplet ground-state suppression factor in \Lead\Lead~collisions at LHC energies, $R_{\Lead\Lead}[\UpsilonMeson[1S]]$, as a function of centrality has emerged, with the initial central temperature $T_0$ and the \UpsilonMeson~formation time~$\FormationTime$ as parameters, in our previous work.
The flat $\TransverseMomentum$-dependence has been explained as a result of the relativistic Doppler effect, which arises whenever the transverse velocity of the quarkonium differs from the expansion velocity of the quark--gluon medium due to the large \UpsilonMeson-meson mass as compared to the parton masses in the plasma~\cite{HoelckNendzigWolschin-2017-PRC}.

However, the results for the first excited \UpsilonMeson[2S] state with the same set of parameters do not show enough suppression in peripheral collisions and hence, additional mechanisms are required.
These should be sufficiently weak to not affect the ground state, but strong enough to have an influence on the excited states.
A possible candidate are electromagnetic interactions, which have so far been neglected in the various quarkonia suppression models.

In particular, the magnetic fields generated in relativistic heavy-ion collisions at the LHC are among the strongest magnetic fields found in nature with estimated field strengths of up to $\SI{5e15}{\tesla}$~\cite{SkokovIllarionovToneev-2009-IJMPA} or, equivalently, $\num{15}\,m_{\Pion}^2/e$ if expressed in terms of the pion mass.
%
The presence of these exceptionally strong fields is expected to give rise to novel phenomena during the collisions, such as chiral charge separation~\cite{KharzeevMcLerranWarringa-2008-NPA} in the emerging quark--gluon plasma (QGP).
Hence it is certainly also possible that they have an effect on the formation and the subsequent in-medium dissociation of excited \Bottomium~states in \Lead\Lead~collisions at the LHC.
It is the purpose of this work to investigate for the first time the small effect of the electromagnetic fields on the dissociation of the \UpsilonMeson[$n$S] states, with emphasis on \UpsilonMeson[2S] since the influence on the ground state will likely be negligible due to its very strong binding.

In the subsequent section, we first estimate the strength of the electromagnetic field generated by two colliding nuclei at the LHC by employing a model that accounts for the conductivity of the quark--gluon medium only~\cite{Tuchin-2013-PRC}.
A brief recap of the medium's hydrodynamic space--time evolution is provided in sect.~\ref{sec:Hydrodynamics}.
We then proceed in sect.~\ref{sec:WaveFunction} to calculate the impact of the electromagnetic field on the \UpsilonMeson-meson wave functions and energies for all states considered, namely, \UpsilonMeson[$n$S] and \chibMeson[$n$P] with $n=1,2,3$, and compute the electromagnetically modified dissociation rates for all these states in sect.~\ref{sec:DecayWidth}.
Results and conclusions are given in the last section.
\section{Electromagnetic fields in relativistic heavy-ion collisions}\label{sec:EMF}
We first investigate the fields generated by a single point-like particle with charge~$q$ traversing a QGP with constant velocity~$v$.
Choosing coordinates $(x^0,x^1,x^2,x^3)$ such that the particle starts at the origin at time $x^0=0$ and moves parallel to the $x^3$-axis, the particle's worldline can be written as
\begin{equation}
	L_q(x^0) = x^0 e_0 + v x^0 e_3
\end{equation}
in terms of the corresponding tetrad $(e_0,e_1,e_2,e_3)$.
The electromagnetic field experienced by a resting observer follows from Maxwell's equations
for an electric charge density
\begin{equation}
	\rho = q \mathinner{\delta(x^1)} \mathinner{\delta(x^2)} \mathinner{\delta(x^3 - v x^0)}
\end{equation}
and electric current density
\begin{equation}
	\vec{j} = v \rho \mathinner{\vec{e}_3} + \Conductivity \PointChargeElectricField
\end{equation}
where $\Conductivity$ is the electrical conductivity of the QGP, $\PointChargeElectricField$ the electric field of the point charge, and $\vec{e}_i$ denotes the projection of $e_i$ to three-dimensional space.
The QGP is assumed to be a linear, homogeneous, and isotropic material with respect to electromagnetic fields and hence to possess a scalar permittivity~$\Permittivity$ and permeability~$\Permeability$.

To date, however, the electromagnetic properties of the QGP~--~especially the dispersion relations of $\Permittivity$ and $\Permeability$~--~are largely unknown, which renders Maxwell's equation impossible to solve.
An approach to circumvent this problem was proposed by Tuchin~\cite{Tuchin-2013-PRC}:
The polarization and magnetization response of the medium are neglected, $\Permittivity = \Permeability = 1$, characterizing it solely by a finite conductivity which increases linearly with its temperature~\cite{Tuchin-2013-AHEP,DingFrancisKaczmarekKarschLaermannEtAl-2011-PRD}
\begin{equation}
	\Conductivity = \SI{5.8}{\MeV} \frac{T}{\QGPFormationTemperature}
	\,,
\end{equation}
where $\QGPFormationTemperature = \SI{160}{\MeV}$ is the critical temperature for QGP formation.
In the relativitic limit $\gamma_q = 1/\sqrt{1 - \smash[b]{v^2}} \gg 1$, this yields~\cite{Tuchin-2013-PRC}
\begin{gather}
	\label{eq:PointChargeElectricField}
	\PointChargeElectricField = \frac{q}{4\CircleConstant} \mathinner{\EulersNumber^{-\RadialDistance^2 \Conductivity / 4 \EffectiveTime}} \left[\frac{\RadialDistance \Conductivity}{2 \EffectiveTime^2} \vec{e}_{\RadialDistance} + \frac{v \RadialDistance^2 \Conductivity / 4 - v \EffectiveTime}{\gamma_q^2 \EffectiveTime^3} \vec{e}_3\right]
	\EquationComma\\
	\label{eq:PointChargeMagneticField}
	\PointChargeMagneticField = \frac{q}{4\CircleConstant} \mathinner{\EulersNumber^{-\RadialDistance^2 \Conductivity / 4 \EffectiveTime}} \frac{v \RadialDistance \Conductivity}{2 \EffectiveTime^2} \vec{e}_{\AzimuthalAngle}
	\EquationComma
\end{gather}
with an effective time~$\EffectiveTime = x^0 - x^3 / v$ and cylindrical coordinates
$\RadialDistance = \sqrt{(x^1)^2 + (x^2)^2}$, $\AzimuthalAngle = \arctan(x^2/x^1)$
which account for the system's symmetries.

Using the linearity of Maxwell's equations, the electric and magnetic fields for a heavy nucleus can be obtained by convolving the point-charge fields with a suitable proton density~$n_Z^\text{con}$~\cite{Tuchin-2013-PRC},
\begin{align}
	\label{eq:NucleusElectricField}
	\NucleusElectricField(x^0,\vec{x}) &= \int\Differential{^3\tilde{x}} \mathinner{n_Z^\text{con}(\vec{\tilde{x}})} \mathinner{\PointChargeElectricField(x^0,\vec{x}-\vec{\tilde{x}})}
	\EquationComma\\
	\label{eq:NucleusMagneticField}
	\NucleusMagneticField(x^0,\vec{x}) &= \int\Differential{^3\tilde{x}} \mathinner{n_Z^\text{con}(\vec{\tilde{x}})} \mathinner{\PointChargeMagneticField(x^0,\vec{x}-\vec{\tilde{x}})}
	\EquationComma
\end{align}
normalized to the nucleus' proton number~$Z$.
The nucleus is assumed to be Lorentz-contracted to a two-dimensional slab in the lab frame due to the high energies at the LHC and Woods-Saxon shaped in its rest frame,
\begin{gather}
	n_Z^\text{con}(\vec{x}) = \delta(x^3) \int\Differential{x^3} n_Z(\vec{x})
	\EquationComma\\
	n_Z(\vec{x}) \propto \left[1 + \exp\!\left(\frac{\Norm{\vec{x}} - R}{a}\right)\right]^{-1}
	\EquationPeriod
\end{gather}
Here, $R$ and $a$ denote the nucleus radius and diffuseness which are given by $R_{\Lead} = \SI{6.62}{\fm}$ and $a_{\Lead} = \SI{0.546}{\fm}$ for lead ions~\cite{VriesJagerVries-1987-ADNDT}.
All nuclear constituents (baryons, partons) are assumed to move at the same longitudinal velocity~$v$.
Due to the form of $n_Z^\text{con}$, the nuclear fields retain some of the symmetries of the point-charge fields~$\PointChargeElectricField,\PointChargeMagneticField$,
\begin{gather}
	\vec{e}_{\AzimuthalAngle} \cdot \NucleusElectricField = \vec{e}_{\RadialDistance} \cdot \NucleusMagneticField = \vec{e}_3 \cdot \NucleusMagneticField = 0
	\EquationComma\\
	\vec{e}_{\RadialDistance} \cdot v \NucleusElectricField = \vec{e}_{\AzimuthalAngle} \cdot \NucleusMagneticField
	\EquationPeriod
\end{gather}

Identifying $x^3$ with the beam direction, $x^0 = 0$ with the moment of the initial collision, and $(x^1,x^3)$ with the reaction plane, the electric field generated by two colliding (identical) nuclei at impact parameter~$\vec{b} = b \mathinner{\vec{e}_1}$ is
\begin{equation}
	\label{eq:TotalElectricField}
	\begin{aligned}
		\TotalElectricField(x^0,\vec{x})
		{}={}&\NucleusElectricField(x^0,\vec{x} + \vec{b} / 2)|_{\vec{v} = +\Abs{v} \vec{e}_3}
		\\
		{}+{}&\NucleusElectricField(x^0,\vec{x} - \vec{b} / 2)|_{\vec{v} = -\Abs{v} \vec{e}_3}
		\EquationPeriod
	\end{aligned}
\end{equation}
The expression for the corresponding magnetic field~$\TotalMagneticField$ follows analogously.

In this simplified approach, the impact of the collision on the structure of the nuclei will mostly be neglected, safe for an instantaneous collective longitudinal deceleration to at least rudimentarily take the effect of baryon stopping~\cite{Bjorken-1983-PRD} into account:
Starting with beam rapidity~$\pm\LongitudinalRapidity_\text{beam}$, the nuclei are slowed down in the moment of the collision to $\LongitudinalRapidity_\text{stopped} < \LongitudinalRapidity_\text{beam}$,
\begin{equation}
	\artanh(\Abs{v}) =
	\begin{cases}
		\LongitudinalRapidity_\text{beam} & \text{for $x^0 < 0$}
		\EquationComma\\
		\LongitudinalRapidity_\text{stopped} & \text{for $x^0 > 0$}
		\EquationPeriod
	\end{cases}
\end{equation}
For $\LongitudinalRapidity_\text{stopped}$, the expected peak rapidity of the participant rapidity distribution is used~\cite{Mehtar-TaniWolschin-2009-PRL,Wolschin-2016-EPJA},
\begin{equation}
	\LongitudinalRapidity_\text{stopped}
	= \frac{1}{1 + \lambda} \left[\LongitudinalRapidity_\text{beam} - \ln\!\big(A^{1/6}\big)\right] + c
	\EquationComma
\end{equation}
which is linked to $\LongitudinalRapidity_\text{beam}$ by the nucleon number~$A$, the saturation-scale exponent~$\lambda = 0.2$ of the gluon saturation momentum, and an empirical constant~$c = -0.2$.

This does not reflect the separation into participants and spectators in the collision for finite impact parameters~$b$.
However, since $\tanh(\LongitudinalRapidity_\text{beam}) \approx 1 \approx \tanh(\LongitudinalRapidity_\text{stopped})$ in \Lead\Lead~collisions at the LHC, the electric and magnetic field of each single constituent in eqs.~(\ref{eq:PointChargeElectricField}) and (\ref{eq:PointChargeMagneticField}) is hardly affected by the slowdown.

Also, even if the direction of the electromagnetic field in close proximity to the nuclei is altered at finite $b$, the overall field strength in the QGP is expected to be reasonably well described by eq.~(\ref{eq:TotalElectricField}) and the corresponding expression for $\TotalMagneticField$.
A more sophisticated treatment of the effect of baryon stopping may in principle be desirable in future revisions of the model, but is unlikely to lead to substantial changes in the results.
\section{Space--time evolution of the thermal quark--gluon medium}\label{sec:Hydrodynamics}
We treat the QGP generated in the relativistic heavy-ion collision as a relativistic perfect fluid of gluons and massless up, down, and strange quarks~\cite{NendzigWolschin-2014-JPGNPP}.
In this first assessment of the importance of electromagnetic field effects, we do not want to include interactions between field and charged medium.
Hence, we use an energy--momentum tensor for the QGP that does not contain any electromagnetic properties,
\begin{equation}
	\label{eq:QGPEnergyMomentumTensor}
	\mathcal{T} = \left(\epsilon + P\right) u \otimes u + P
	\,,
\end{equation}
but is composed solely of the fluid's internal energy density $\epsilon$, pressure $P$, and four-velocity $u$.
The equation of state is chosen appropriately for a perfect relativistic fluid
\begin{equation}
	\label{eq:QGPEquationOfState}
	P = c_\text{s}^2 \epsilon
	\,,\qquad
	c_\text{s} = \frac{1}{\sqrt{3}}
	\,,\qquad
	\epsilon = \epsilon_0 T^4
	\,.
\end{equation}
Imposing four-momentum conservation, $\nabla\cdot\mathcal{T} = 0$, we obtain the general equations of motion
\begin{equation}
	\label{eq:EnergyMomentumConservation}
	\frac{1}{\sqrt{\Abs{\det g}}} \partial_\mu \left(\sqrt{\Abs{\det g}} \mathcal{T}\indices{^\mu_\alpha}\right) = \frac{1}{2} \mathcal{T}^{\mu\nu} \partial_\alpha g_{\mu\nu}
\end{equation}
governing the medium's hydrodynamical expansion in a frame of reference with space-time metric~$g$.

We evaluate these equations in the longitudinally co-moving frame $(\ProperTime,x^1,x^2,\LongitudinalRapidity)$ which is linked to the laboratory frame introduced in sect.~\ref{sec:EMF} via
\begin{equation}
	\ProperTime = \sqrt{(x^0)^2 - (x^3)^2}
	\,,\qquad
	\LongitudinalRapidity = \artanh(x^3\!/x^0)
	\,.
\end{equation}
In this frame, the $\LongitudinalRapidity$-component of the fluid velocity vanishes, $e_{\LongitudinalRapidity} \cdot u = 0$, while the metric takes the form
\begin{equation}
	\label{eq:LCFMetric}
	g = -\Differential{\ProperTime}^2 + \ProperTime^2 \Differential{\LongitudinalRapidity}^2 + (\Differential{x^1})^2 + (\Differential{x^2})^2
	\,.
\end{equation}
Inserting eqs.~(\ref{eq:QGPEnergyMomentumTensor}), (\ref{eq:QGPEquationOfState}), and (\ref{eq:LCFMetric}) into eqs.~(\ref{eq:EnergyMomentumConservation}) yields
\begin{equation}
	\label{eq:QGPEquationsOfMotion}
	\partial_\mu \left(\ProperTime T^4 u^\mu u_\alpha\right) = -\frac{\ProperTime}{4} \partial_\alpha T^4
	\,,\qquad
	\partial_\mu \left(\ProperTime T^3 u^\mu\right) = 0
	\,,
\end{equation}
where the second part of the equation follows from a projection of eqs.~(\ref{eq:EnergyMomentumConservation}) onto the fluid velocity~$u$.

Eqs.~(\ref{eq:QGPEquationsOfMotion}) are solved numerically, beginning after the passing of an initial local thermalization time~$\ThermalizationTime = \SI{0.1}{\fm}$ in the longitudinally co-moving frame.
For a relativistic heavy-ion collision at the LHC, the initial conditions in the transverse plane $(x^1,x^2)$ are chosen as follows:
\begin{gather}
	u^1(\ThermalizationTime) = u^2(\ThermalizationTime) = 0
	\,,\\
	T(b,\ThermalizationTime,x^1,x^2) = T_0 \sqrt[3]{\frac{N_\text{mix}(b,x^1,x^2)}{N_\text{mix}(0,0,0)}}
	\,,\\
	N_\text{mix} = f \NumberOfParticipants + (1-f) \NumberOfCollisions
	\,,\quad
	f = \num{0.8}
	\,,
\end{gather}
where $\NumberOfCollisions$ is the number of binary collisions, $\NumberOfParticipants$ the number of participating nucleons, and $N_\text{mix}$ the number of particle-producing sources~\cite{ALICE-2011-PRL}.
The initial central temperature~$T_0$ is determined through a fit to the $\TransverseMomentum$-dependent centrality-integrated experimental $R_{AA}$ data for the \UpsilonMeson[1S] state.
For a \Lead\Lead~collision at $\CMEPerNucleonPair = \SI{2.76}{\TeV}$, the value $T_0 \approx \SI{480}{\MeV}$ is obtained using a formation time~$\FormationTime = \SI{0.4}{\fm}$ for all six states involved.
These parameters are unchanged as compared to our previous work \cite{HoelckNendzigWolschin-2017-PRC}, where they had been updated from \cite{NendzigWolschin-2014-JPGNPP} due to more refined treatment of the relativistic Doppler effect.

\begin{figure}
	\centering
	\includegraphics[scale=0.8]{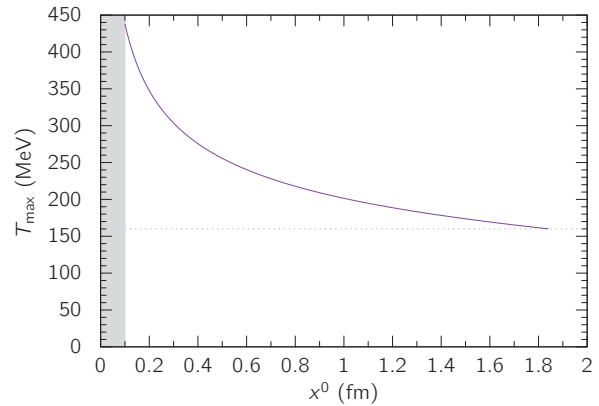}
	\caption{\label{fig:Temperature}
		Maximum medium temperature in a \Lead\Lead~collision with $\CMEPerNucleonPair = \SI{2.76}{\TeV}$ and $b = \SI{7}{\fm}$ as a function of the elapsed time~$x^0$.
		The shaded area indicates the local thermalization period following the initial collision.
		The critical temperature~$\QGPFormationTemperature$ for QGP formation is marked by a dotted line.
	}
\end{figure}

Fig.~\ref{fig:Temperature} shows the time evolution of the medium's maximum temperature.
The QGP lifetime implied by the diagram of $\SI{\sim 2}{\fm}$ is about one fifth of the value experiment suggests~\cite{ALICE-2011-PLB}.
This is presumably a consequence of our assumption of an ideal fluid which results in faster expansion and cooling of the medium due to its lack of viscosity.
\section{Impact on the \UpsilonMeson-meson wave function}\label{sec:WaveFunction}
In the current model, the potential non-relativistic QCD (pNRQCD) formalism~\cite{CaswellLepage-1986-PLB,BrambillaGhiglieriVairoPetreczky-2008-PRD} is used to describe \UpsilonMeson~mesons embedded in a thermal medium:
The radial wave functions~$g_{nl}$ and the corresponding energies~$E_{nl}$ and damping decay widths~$\Gamma_{\text{damp},nl}$ for the six \UpsilonMeson~states\footnote{In the following, the term ``\UpsilonMeson~state'' will be used for all \Bottomium~mesons with spin $s=1$ to ease notation.} considered in the model~--~\UpsilonMeson[1S], \UpsilonMeson[2S], \UpsilonMeson[3S], and \chibMeson[1P], \chibMeson[2P], \chibMeson[3P]~--~are obtained by solving the radial time-independent Schr\"odinger equation for a complex potential~$V_{nl}$~\cite{NendzigWolschin-2014-JPGNPP,HoelckNendzigWolschin-2017-PRC},
\begin{equation}
	\label{eq:SchroedingerEquation}
	\partial^2_\Displacement g_{nl} = m_{\BottomQuark} \left(V_{nl} - E_{nl} + \tfrac{\ImaginaryUnit}{2} \Gamma_{\text{damp},nl}\right) g_{nl}
	\EquationComma
\end{equation}
where $\Displacement = \Norm{\vec{\Displacement}}$ measures the displacement of the bottom and antibottom quark, $m_{\BottomQuark}$ is the bottom-quark mass, and $n\in\{1+l,2+l,3+l\}$ and $l\in\{0,1\}$ denote the principal and angular-momentum quantum number, respectively.
Since $V_{nl}$ is implicitly dependent on $E_{nl}$ due to the energy-dependence of the contained strong coupling, we solve eq.~(\ref{eq:SchroedingerEquation}) numerically by means of an iterative scheme.

Including an external electromagnetic field $\vec{E},\vec{B}$ into the pNRQCD formalism to first order gives rise to two additional terms in the complex potential,
\begin{equation}
	\label{eq:ElectromagneticPotentialTerms}
	V_{nl} = V_{nl}|_{\vec{E}=\vec{B}=\vec{0}} - \vec{p} \cdot \vec{E} - \vec{\mu} \cdot \vec{B}
	\EquationPeriod
\end{equation}
The symbols
$\vec{p} = q_{\BottomQuark} \vec{\Displacement}$ and $\vec{\mu} = q_{\BottomQuark} \vec{\sigma} / m_{\BottomQuark}$ denote the electric and magnetic dipole moment of the \UpsilonMeson~state, given in terms of the bottom-quark charge~$q_{\BottomQuark} = -e/3$ and the spin operator~$\vec{\sigma}$.
The magnetic term in eq.~(\ref{eq:ElectromagneticPotentialTerms}) acts on the spin part of the \UpsilonMeson-meson wave function and leads to a shift~$\Delta_{nl}$ in the state's energy~\cite{Filip-2013-PoS,AlfordStrickland-2013-PRD},
\begin{equation}
	E_{nl} = E_{nl}|_{\vec{B}=\vec{0}} - (-1)^s \delta_{0 m_s} \Delta_{nl}
	\EquationPeriod
\end{equation}
The shift only occurs for spin projection quantum number~$m_s=0$ with a sign depending on the state's spin quantum number~$s\in\{0,1\}$.
Its magnitude
\begin{gather}
	\Delta_{nl} = \frac{\varepsilon_{nl}}{2} \left(\sqrt{1 + \chi_{nl}^2} - 1\right)
	\EquationComma\quad
	\chi_{nl} = \frac{4 \mu_{\BottomQuark} \Norm{\vec{B}}}{\varepsilon_{nl}}
\end{gather}
is subject to the quark magneton~$\mu_{\BottomQuark} = q_{\BottomQuark}/(2 m_{\BottomQuark})$ and the energy gap~$\varepsilon_{nl}$ between the $s=0$ and $s=1$ state.
For the states with $n=1+l$ and $n=2+l$, $\varepsilon_{nl}$ can be determined from the difference between the invariant masses, $\varepsilon_{nl} = m_{nl}|_{s=1} - m_{nl}|_{s=0}$, cf.~table~\ref{tab:HyperfineSplitting}.
For $n=3+l$, the magnetic energy shift is neglected due to a lack of data for these states.

\begin{table}
	\centering
	\caption{\label{tab:HyperfineSplitting}
		Invariant mass difference between the $s=1$ and $s=0$ \Bottomium~state for $n\in\{1+l,2+l\}$ and $l\in\{0,1\}$~\cite{PDG-2016-CPC}.
		The symbol $j|_{s=1}$ indicates the total angular momentum quantum number of the $s=1$ state.
	}
	\def\arraystretch{1.5}
	\begin{tabular}{ccrr}
		\hline\hline
		\multicolumn{2}{c}{$m_{nl}|_{s=1} - m_{nl}|_{s=0}$ (\si{\MeV})} & \multicolumn{1}{c}{$n-l=1$} & \multicolumn{1}{c}{$n-l=2$} \\
		\hline
		$l=0$ & $j|_{s=1}=1$ & \num{+62.3+-3.2} & \num{+24+-4} \\
		\hline
		& $j|_{s=1}=0$ & \num{-39.9+-1.4} & \num{-27.3+-1.9} \\
		$l=1$ & $j|_{s=1}=1$ & \num{-6.5+-1.2} & \num{-4.3+-1.8} \\
		& $j|_{s=1}=2$ & \num{+12.9+-1.2} & \num{+8.9+-1.8} \\
		\hline\hline
	\end{tabular}
\end{table}

\begin{figure}
	\centering
	\includegraphics[scale=0.8]{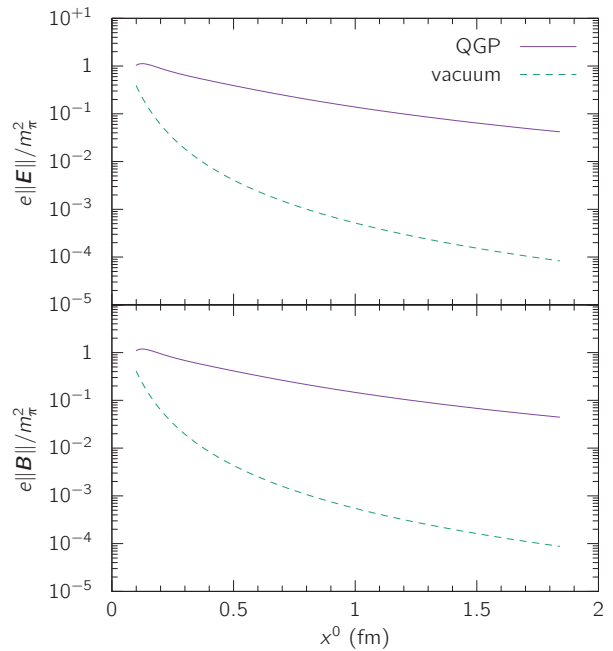}
	\caption{\label{fig:EMF}
		Electric (top) and magnetic (bottom) field strength at position~$\vec{x}=\num{0.5}R_{\Lead}\vec{e}_2$ generated in a \Lead\Lead~collision with $\CMEPerNucleonPair = \SI{2.76}{\TeV}$ and $b = \SI{7}{\fm}$ as a function of elapsed time~$x^0$.
		The solid lines indicate the field strength in presence of a QGP; for comparison, the vacuum solution is shown as dashed lines.
		The finite conductivity of the QGP sustains the electromagnetic field, but counteracts its propagation, which is visible at small times $x^0$.
		See also fig.~2 in \cite{GursoyKharzeevRajagopal-2014-PRC} for a similar calculation.
	}
\end{figure}

Employing the formalism introduced in sect.~\ref{sec:EMF} to calculate the electromagnetic field of two colliding \Lead~nuclei with $\CMEPerNucleonPair = \SI{2.76}{\TeV}$ at the LHC reveals that the field strength should satisfy $\Norm{\vec{E}},\Norm{\vec{B}} \lesssim m_{\Pion}^2/e$ in the rest frame of produced \UpsilonMeson~mesons.
This is a consequence of the nearly exponential decay of the field strength with time, as can be seen in fig.~\ref{fig:EMF}, and the \UpsilonMeson~meson's finite formation time; the latter is set to $\FormationTime = \SI{0.4}{\fm}$ in the model for all six \UpsilonMeson~states considered.

Although the presence of the QGP significantly prolongs the lifetime of the electromagnetic field, it still drops by an order of magnitude between the thermalization of the QGP at $\SI{\sim 0.1}{\fm}$ and the formation of the first \UpsilonMeson~states.
As a consequence, the electric part of $V_{nl}$ is considerably smaller than the leading Cornell-like contribution from the strong force,
\begin{equation}
	\label{eq:ElectricDipoleTermEstimate}
	\frac{\Abs{\vec{p}\cdot\vec{E}}}{\sigma \Displacement}
	\lesssim \frac{m_{\Pion}^2/3}{\sigma}
	\lesssim \num{0.04}
\end{equation}
with string tension~$\sigma = \SI{0.192}{\GeV\squared}$, and hence, the electric term can be simplified in good approximation by replacing the running displacement variable~$r$ with its root-mean-square~$\DisplacementRMS$, thereby fixing the electric dipole moment's magnitude for each state,
$\Norm{\vec{p}} \mapsto \Abs{q_{\BottomQuark}} \DisplacementRMS$.
The root-mean-square can be evaluated neglecting the electric field term
\begin{equation}
	\DisplacementRMS = \left[\int_{0}^{\infty}\Differential{\Displacement} \Displacement^2 \Abs{g_{nl}}^2\right]^{1/2}_{\vec{E}=\vec{0}}
\end{equation}
due to the electric term's minor importance for the wave function.
Its numerical value lies in the range $\SI{0.2}{\fm} \lesssim \DisplacementRMS \lesssim \SI{1.0}{\fm}$, being smaller at low temperatures of the surrounding QGP or for states close to the ground state and larger at high temperatures or for excited states.

Comparing the magnetic energy shift~$\Delta_{nl}$ to the leading Cornell-like string term,
\begin{equation}
	\label{eq:MagneticDipoleTermEstimate}
	\frac{\Delta_{nl}}{\sigma \Displacement}
	\lesssim \frac{m_{\Pion}^2/(3 m_{\BottomQuark})}{\sigma \DisplacementRMS}
	\lesssim \num{0.008}
	\EquationComma
\end{equation}
it becomes apparent that the magnetic part of $V_{nl}$ is even less important than its electrical counterpart.
This is ultimately a consequence of the fact that the meson's effective magnetic dipole moment $2\mu_{\BottomQuark}$ is an order of magnitude smaller than its electric dipole moment, $2\Abs{\mu_{\BottomQuark}} / \Norm{\vec{p}} \lesssim \num{0.2}$, while the electric and magnetic field in the collision are of similar strength.

The shape of the \UpsilonMeson-meson wave functions should therefore be virtually unaffected by the electromagnetic field generated in heavy-ion collisions at LHC energies.
Consequently, the electromagnetic terms in $V_{nl}$ are ignored in the computation of $g_{nl}$ and solely treated as a perturbation of the \UpsilonMeson~meson's energy
\begin{multline}
	\label{eq:ElectromagneticEnergyShift}
	E_{nl}
	= E_{nl}|_{\vec{E}=\vec{B}=\vec{0}}
	\\
	- q_{\BottomQuark} \DisplacementRMS \Norm{\vec{E}} \cos(\ElectricDipoleAlignment)
	\\
	+ \delta_{0 m_s} \Delta_{nl}(\Norm{\vec{B}})
\end{multline}
where $\ElectricDipoleAlignment = \measuredangle(\vec{p},\vec{E})$ measures the alignment of the electric dipole to the electric field.
%
\section{Electromagnetic dissociation of \UpsilonMeson~states}\label{sec:DecayWidth}
Two primary sources for \UpsilonMeson~meson dissociation are considered in the current model:
The imaginary part of the complex potential~$V_{nl}$ gives rise to a decay width~$\Gamma_{\text{damp},nl}$ accounting for the Landau damping of the \Bottomium~binding.
In addition, a gluo-dissociation decay width~$\Gamma_{\text{gluo},nl}$ which reflects gluon-induced transitions from bound color-singlet to unbound color-octet states is calculated through a convolution of the color-singlet-to-octet dipole transition cross section with the thermal gluon distribution~\cite{BrezinskiWolschin-2012-PLB,NendzigWolschin-2014-JPGNPP}.
The damping and gluo-dissociation processes act on different energy scales~\cite{BrambillaEscobedoGhiglieriVairo-2011-JHEP} and are therefore treated individually in accordance with the separation of scales in pNRQCD; the combined decay width experienced by \UpsilonMeson~mesons in the QGP is given by the incoherent sum $\Gamma_{\text{damp},nl} + \Gamma_{\text{gluo},nl}$.

\begin{figure}
	\centering
	\includegraphics[scale=0.8]{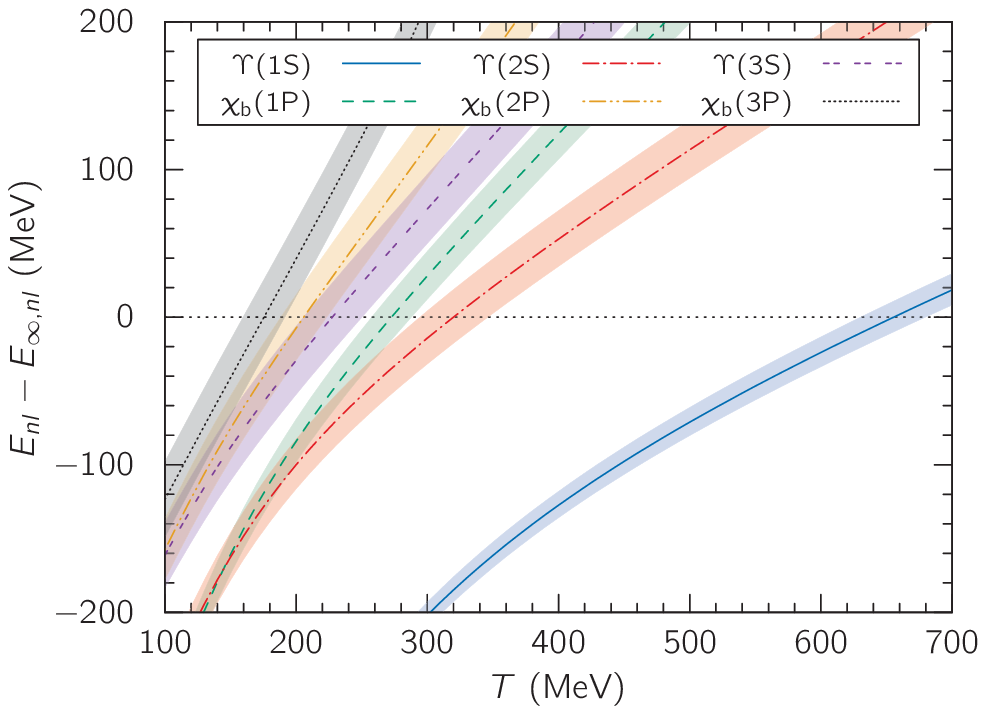}
	\caption{\label{fig:Energy}
		Meson energies as a function of the medium temperature.
		Lines indicate the energy values in the absence of electromagnetic fields; the shaded areas indicate the deviation from these values for an electromagnetic field strength typically encountered during LHC collisions $e\Norm{\vec{E}}/m_{\Pion}^2 = e\Norm{\vec{B}}/m_{\Pion}^2 = 1$ and all possible electric dipole alignment angles~$\ElectricDipoleAlignment \in [0,\CircleConstant]$.
		The dissociation temperatures $T_{\text{diss},nl}$ are implicitly defined by the intersections with the dotted line at $E_{nl} - E_{\infty,nl} = 0$.
	}
\end{figure}

A third, indirect contribution comes from screening: 
Bound states can only exist for energies $E_{nl}$ below the continuum threshold~$E_{\infty,nl} = \lim_{r\to\infty} \Re V_{nl}$; if a meson possesses a higher energy, it dissolves into an unbound \Bottomium~pair.
Combining all three sources yields the total in-medium decay width
\begin{equation}
	\label{eq:TotalDecayWidth}
	\Gamma_{\text{tot},nl} =
	\begin{cases}
		\Gamma_{\text{damp},nl} + \Gamma_{\text{gluo},nl} & \text{for } E_{nl} < E_{\infty,nl}
		\EquationComma\\
		\infty & \text{for } E_{nl} > E_{\infty,nl}
		\EquationPeriod
	\end{cases}
\end{equation}
In this approach to \UpsilonMeson-meson dissociation, the energy difference $E_{nl} - E_{\infty,nl}$ implicitly defines a dissociation temperature $T_{\text{diss},nl}$ above which the meson spontaneously decays.
Due to the electromagnetic contribution to $E_{nl}$, $T_{\text{diss},nl}$ is altered in the presence of an electromagnetic field, cf.~fig.~\ref{fig:Energy}.

The sign and magnitude of the electric energy shift in eq.~(\ref{eq:ElectromagneticEnergyShift}) depend on the angle $\ElectricDipoleAlignment = \measuredangle(\vec{p},\vec{E})$, the orientation of the electric dipole moment to the electric field.
An increased $E_{nl}$ monotonically translates into a smaller $T_{\text{diss},nl}$ which in turn results in increased suppression of the associated \UpsilonMeson~state.
Hence, the range of results spanned by the two limiting cases $\ElectricDipoleAlignment \in \{0,\CircleConstant\}$, corresponding to total (anti-)parallel alignment of all dipoles with the field,
\begin{equation}
	E_{nl}^{\text{max/min}}
	= E_{nl}|_{\vec{E}=\vec{0}}
	\mp q_{\BottomQuark} \DisplacementRMS \Norm{\vec{E}}
	\EquationComma
\end{equation}
necessarily contains all possible outcomes.
This can be used to significantly reduce the computing effort required.

Eventually, the pre-cascade and final nuclear modification factors~$\PreCascadeNuclearModificationFactor$ and $\NuclearModificationFactor$, respectively, are calculated from $\Gamma_{\text{tot},nl}$ as outlined in \cite{HoelckNendzigWolschin-2017-PRC}.
To reflect the relativistic Doppler effect in the QGP, the angular-averaging procedure described therein is used.
\section{Results and conclusion}
To assess the importance of the electromagnetic interactions presented in sects.~\ref{sec:WaveFunction} and \ref{sec:DecayWidth}, the nuclear modification factors are calculated both with and without taking the electromagnetic field generated by the colliding nuclei into account.
The latter case will be referred to in short as the ``field-free scenario'' in the following discussion.

\begin{figure*}
	\centering
	\includegraphics[scale=0.8]{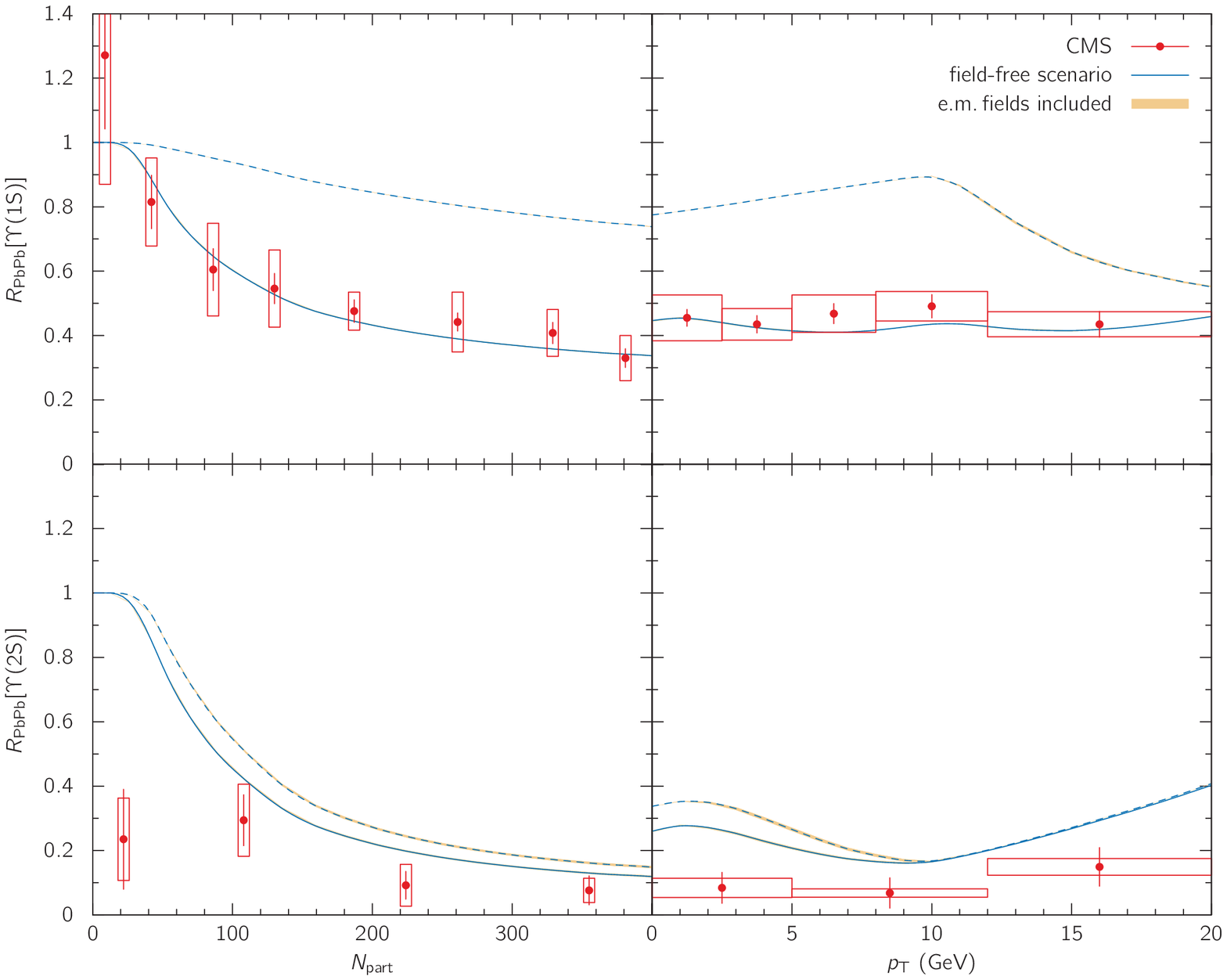}
	\caption{\label{fig:RAA}
		Nuclear modification factors for \UpsilonMeson[1S] as a function of the number of participants~$\NumberOfParticipants$ (top left) and transverse momentum~$\TransverseMomentum$ (top right) as well as for \UpsilonMeson[2S] as a function of $\NumberOfParticipants$ (bottom left) and $\TransverseMomentum$ (bottom right) in a \Lead\Lead~collision at $\CMEPerNucleonPair = \SI{2.76}{\TeV}$.
		The computed final nuclear modification factor~$\NuclearModificationFactor$ after the decay cascade and pre-cascade nuclear modification factor~$\PreCascadeNuclearModificationFactor$ in the ``field-free scenario'' are depicted as solid and dashed lines, respectively.
		If the electromagnetic field is taken into account, $\PreCascadeNuclearModificationFactor$ and $\NuclearModificationFactor$ deviate from the ``field-free scenario'' as indicated by the shaded areas surrounding the solid and dashed curves.
		Extent and sign of the deviation are subject to the electric dipole alignment~$\ElectricDipoleAlignment \in [0,\CircleConstant]$ of each single meson; the shaded areas cover all possible configurations.
		Data from~\cite{CMS-2017-PLB} are shown for comparison.
	}
\end{figure*}

Fig.~\ref{fig:RAA} shows our computed results for the nuclear modification factors $\PreCascadeNuclearModificationFactor$ and $\NuclearModificationFactor$ for the \UpsilonMeson[1S] and \UpsilonMeson[2S] states.
Clearly, the impact of the electromagnetic field generated by the colliding nuclei on \UpsilonMeson~meson suppression via screening is negligible:
Using the shorthand
\begin{equation}
	\Increment\NuclearModificationFactor = \left\lvert\NuclearModificationFactor - \NuclearModificationFactor^{\vec{E}=\vec{B}=\vec{0}}\right\rvert
	,
\end{equation}
the observed deviation of the nuclear modification factors from the ``field-free scenario'' is $\Increment\NuclearModificationFactor \lesssim \num{2e-3}$ for the \UpsilonMeson[1S] and thus significantly less than $\SI{1}{\percent}$.
While this is not surprising in case of the relatively stable $s=1$ ground state, the much more volatile \UpsilonMeson[2S] state is seen to be equally unaffected, $\Increment\NuclearModificationFactor \lesssim \num{3e-3}$, which points to a general insignificance of this mode of suppression.
If only the impact of the magnetic field is considered, the effect is even smaller, $\Increment\NuclearModificationFactor|_{\vec{E}=\vec{0}} \lesssim \num{e-4}$, which is consistent with our estimates from eqs.~(\ref{eq:ElectricDipoleTermEstimate}) and (\ref{eq:MagneticDipoleTermEstimate}).


In summary, the electromagnetic field effects on the dissociation of \UpsilonMeson~mesons studied in this work have proven to be negligible in \Lead\Lead~collisions at LHC-Run-I energies.
Since the electromagnetic field strength depends only very weakly on rapidity, the same conclusion applies to the recent LHC-Run-II where the center-of-mass energy was increased to $\CMEPerNucleonPair = \SI{5.02}{\TeV}$ (not shown here).

In fact, according to our tests, it would be necessary to increase the field strength by a factor of \num{10} in order to obtain a noticeable deviation from the field-free scenario ($\Increment\NuclearModificationFactor \lesssim \num{1.5e-2}$ for \UpsilonMeson[1S], $\Increment\NuclearModificationFactor \lesssim \num{3e-2}$ for \UpsilonMeson[2S]).
Even when considering the inaccuracies of our current model, this would presumably require the use of significantly heavier nuclei which is not experimentally feasible.
Improvements in the theoretical description of the medium's conductivity do not affect the maximum strength of the electromagnetic field, but only its lifetime which barely affects the final $R_{AA}$ results.

It may, however, be worthwhile to investigate the effect of the short-lived electromagnetic field on the formation of \UpsilonMeson~mesons in the medium and on the corresponding formation times:
The strong electromagnetic field in the early stages of the collision could inhibit or foster the formation process of individual \UpsilonMeson~states leading to a change in the initial meson populations in \Lead\Lead~collisions which does not occur in \Proton\Proton, where the field is considerably weaker.

\begin{acknowledgement}
	We thank F.~Nendzig (now at Accso~--~Accelerated Solutions GmbH, Darmstadt) for providing the C\#~program to compute \UpsilonMeson~suppression in strong fields that was used as a basis to calculate the electromagnetic field effects.
\end{acknowledgement}
\bibliographystyle{epj}
\bibliography{hw_epja17.bib}
\end{document}